\documentclass[%
pra, 
 mph,%
 amsmath,amssymb,
 reprint,%
]{revtex4-2}

\usepackage{graphicx}
\usepackage{dcolumn}
\usepackage{bm}
\usepackage{hyperref}
\usepackage[caption=false]{subfig}

\usepackage[mathlines]{lineno}
\modulolinenumbers[5]
\linenumbers\relax 

\begin{document}
\nolinenumbers
\preprint{AAPM/123-QED}

\title{Above-threshold ionization at laser intensity greater than $10^{20}$ W/cm$^{2}$}

\author{A. Yandow}
\affiliation{Center for High Energy Density Science, The University of Texas at Austin, 2515 Speeday Stop C1600, Austin, TX 78712}%
\affiliation{Lawrence Livermore National Laboratory, Livermore, CA 94551}
\email{yandow1@llnl.gov}
\author{T. N. Ha}%
\author{C. Aniculaesei}
\author{H. L. Smith}
\author{C. G. Richmond}
\author{M. M. Spinks}
\author{H. J. Quevedo}
\author{S. Bruce}
\author{M. Darilek}
\author{C. Chang}
\author{D. A. Garcia}
\author{E. Gaul}
\author{M. E. Donovan}
\author{B. M. Hegelich}
\author{T. Ditmire}
\affiliation{Center for High Energy Density Science, The University of Texas at Austin, 2515 Speedway Stop C1600, Austin, TX 78712}%

\date{\today}

\begin{abstract}
We present the first experimental observation of above-threshold ionization (ATI) electrons produced by ionization of the neon K-shell in a laser field where intensity exceeds 10$^{20}$ W/cm$^{2}$. An array of plastic scintillating calorimeter detectors was used to measure the high-energy electrons at four angles in the laser forward direction. Coarse energy resolution was obtained using aluminum filters of several thicknesses to block lower-energy electrons. A threshold intensity around $2 \times 10^{20}$ W/cm$^{2}$ is observed for production of energetic ATI electrons in the laser forward direction, with maximum electron energy exceeding 10 MeV. L-shell electrons with energies $<1.4$ MeV are scattered further forward along the laser direction than expected. We present comparisons of the measured total electron energies to the predictions of a Monte Carlo models employing the ADK-PPT ionization model and the Augst barrier suppression ionization model. 
\end{abstract}

\maketitle
\section{Introduction}
        Above-threshold ionization (ATI) is the process by which an electron absorbs many more photons than required for ionization during a laser-atom interaction. Absorption of a single additional photon over the required threshold was observed in 1979 by Agostini \textit{et al}. \cite{Agostini1979a}. The modern two-step model of ATI was proposed by Corkum \textit{et al}. to explain the absorption of thousands of laser photons above the threshold \cite{Corkum1989a}, and it has been extended to explain high-harmonic generation in gases \cite{Corkum1993}\cite{Krause1992} and nonsequential double ionization (NSDI)\cite{Watson1997}\cite{Walker1994}\cite{Fittinghoff1992a}. The two-step model of ATI in a strong, near-infrared laser field describes the ionization process as a quantum mechanical tunneling model and predicts the subsequent electron dynamics by integrating the classical Lorentz force equations. The two-step model of ATI has explained a number of experimental observations, including relativistic electrons gaining a momentum component in the laser forward direction\cite{Moore1995} and the preferential ejection of ATI electrons along the laser polarization direction \cite{McNaught1998}\cite{Dichiara2008}. At present, ATI electrons with energies exceeding 1 MeV from argon and xenon have been observed, corresponding to the absorption of one million excess photons above the ionization threshold\cite{Dichiara2008}\cite{Ekanayake2013}. 
        
        Tunneling ionization becomes the dominant ionization mechanism for near-infrared laser fields when intensity exceeds $10^{14}$ W/cm$^{2}$. Tunneling ionization first observed in the pioneering experiments of Augst \textit{et al}. \cite{Augst1989a}\cite{Augst1991}. The Ammosov-Krainov-Delone and Perelomov-Popov-Terent'ev (ADK-PPT) model of tunneling ionization \cite{Perelomov1966}\cite{Ammosov1986} has been verified with precision measurements of argon charge states at intensity exceeding $2 \times 10^{19}$ W/cm$^{2}$ \cite{Chowdhury2001}. The highest ion charge states observed experimentally are Ar$^{16+}$\cite{Chowdhury2001}\cite{Yamakawa2003}, Xe$^{26+}$\cite{Yamakawa2003}\cite{Akahane2006a}, and Kr$^{24+}$\cite{Akahane2006a}. The probability of tunneling ionization is a strongly nonlinear function of laser intensity, leading to the use of high ion charge states as a direct peak laser intensity diagnostic with varying degrees of success \cite{Akahane2006a}\cite{Link2006}. With laser intensity estimates calculated from indirect diagnostics exceeding $2 \times 10^{22}$ W/cm$^{2}$ \cite{Tiwari2019}\cite{Yoon2019} and 10-PW-class laser facilities in their final stages of development\cite{Rus2017}, there has been considerable renewed interest in highly-charged ions as a direct peak laser intensity diagnostic. Recent numerical studies of ionization have developed a tunneling cascade ionization model for complex ions in a laser field \cite{Ciappina2019}, demonstrated that K-shell ionization yields are the most robust when considering different ionization models \cite{Ciappina2020}, and identified features of the ionization yield curves that are robust when considering different intensity distributions at the focal plane \cite{Ciappina2020a}. Monte Carlo simulations of ionization that include ion motion in the laser field demonstrate the ions can be accelerated to energies that make conventional time-of-flight ion yield measurements impossible at intensity above $10^{21}$ W/cm$^{2}$ \cite{Yandow2019}, so we explore the  detection of the ATI electrons from these high charge states for future ionization physics experiments and direct laser intensity diagnostics. 
        
        Modulations in ATI electron energy spectra and angular distributions corresponding to ionization of different electron shells of the target atomic species has been observed for the N, M, and L-shells of krypton and xenon \cite{Ekanayake2013}. These modulations arise from the large difference in ionization potential between the atomic shells, with ATI electrons produced by ionization of a deeply-bound state accelerated to higher energies by a stronger laser electric field than an outer shell state\cite{Ekanayake2013}. The large ionization potential gap between the L-shell and K-shell of neon will result in a strong modulation of the energy spectrum and angular distribution, with the K-shell electrons gaining about an order of magnitude more energy than the L-shell electrons. This strong modulation in both the energy spectrum and the angular distribution raises the possibility of a novel direct laser intensity diagnostic, where the production of highly-charged ion states can be inferred by the detection of high-energy ATI electrons ejected in the laser forward direction. Such a diagnostic will be relatively easy to execute experimentally using a low-density stream of noble gas as a target and a magnetic spectrometer to detect the ATI electrons. The ionization of the K-shell in noble gases would allow for direct laser intensity measurement around 10$^{20}$ W/cm$^{2}$ (Ne$^{9+}$), $3 \times 10^{21}$ W/cm$^{2}$ (Ar$^{17+}$), and $10^{23}$ W/cm$^{2}$ (Kr$^{35+}$) in any laser field, provided the ADK-PPT tunneling ionization model can be verified experimentally on a laser system with reliable indirect diagnostics at these intensities as well. 
        
        ATI electrons hold significant promise as a direct laser intensity diagnostic between $10^{20}$ and $10^{24}$ W/cm$^{2}$, where ponderomotive expulsion of the ions becomes a significant obstacle to direct ionization yield measurement\cite{Yandow2019}. Recently proposed techniques using vacuum-accelerated electrons and protons \cite{Vais2020}\cite{Vais2020c} will not yield an accurate intensity estimate without a well-known pulse duration when prepulses scatter target electrons \cite{Vais2018} and the ions gain only a fraction of their ponderomotive potential energy \cite{Vais2020c}. The strong nonlinearity of the tunneling ionization rate prevents the K-shell electrons from being scattered by prepulses even when laser contrast is as low as $10^{-3}$. The ceiling intensity of our method is about $10^{24}$ W/cm$^{2}$, above which expulsion of highly-charged ions before the arrival of the peak laser field strength would prevent K-shell ionization \cite{Yandow2019} and an ion ponderomotive diagnostic such as that proposed in \cite{Vais2020} would be most appropriate. 
        
        We report in this paper the observation of multi-MeV ATI electrons produced by the interaction of a low-density neon gas jet ($< 3 \times 10^{13}$ cm$^{-3}$) with a well-characterized laser pulse with intensity exceeding $10^{20}$ W/cm$^{2}$. We observe these ATI electrons on four plastic scintillating calorimeter detectors positioned in the laser forward direction and along the plane of laser polarization. We compare the observed integrated ATI electron energy yields to the predictions of an ADK-PPT Monte Carlo model of ATI and a barrier suppression model of ATI, using methods similar to those described previously elsewhere\cite{Yandow2019}. The measurements have qualitative similarities with the models' predictions, including the existence of a threshold intensity above which the ionization probability increases rapidly with intensity and a saturation intensity above which the ATI electron energy yield is dominated by the focal volume rather than by the probability of ionization in the center of the focus. However, we observe poor quantitative agreement with the modeling, which significantly underestimates the observed threshold intensity by a factor between two and three. We also observe electrons with energies exceeding $10$ MeV for the first time \cite{YandowPRL2023}. 
        
\section{Experimental Design}

\begin{figure}[t!]
\includegraphics[width = \linewidth]{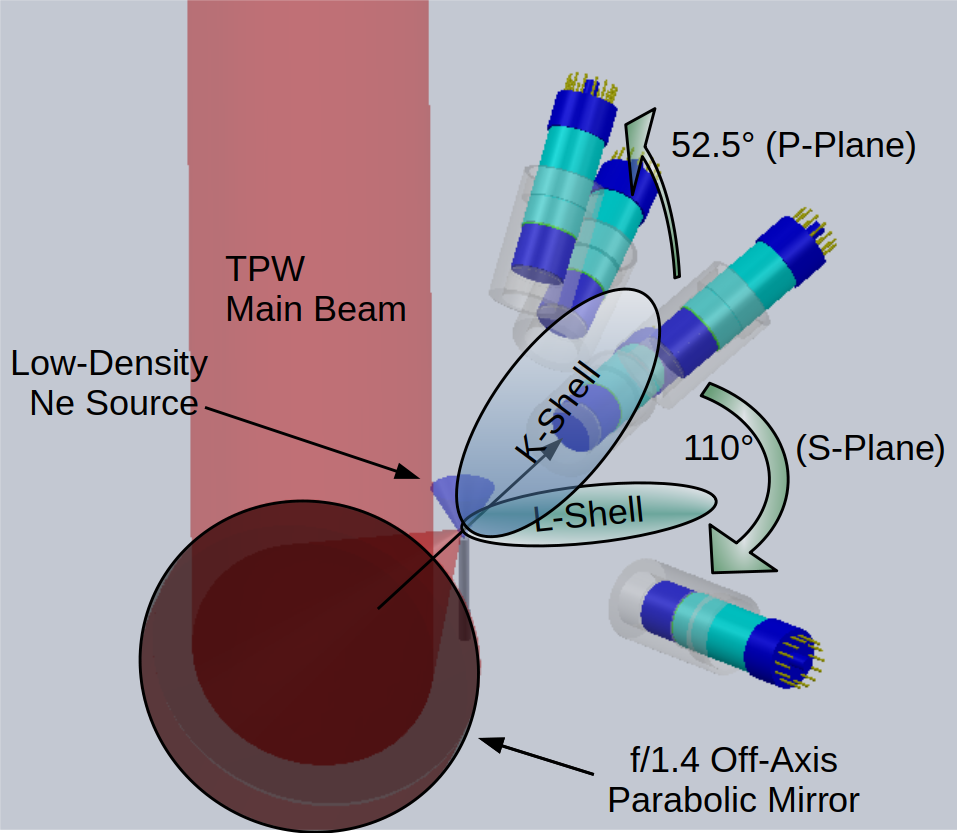}
\caption{\label{fig1} Diagram of the experimental setup, showing five detectors arranged around the laser-atom interaction region. Three detectors were placed in the plane of laser polarization at angles of $0^{\circ}$, $30^{\circ}$, and $53^{\circ}$ from the laser forward direction. One additional detector was placed $60^{\circ}$ out of the plane of polarization and $43^{\circ}$ from the laser forward direction. A control detector was placed $110^{\circ}$ from the forward direction out of the polarization plane.}
\end{figure}

A diagram of the experimental setup is given in Fig. \ref{fig1}.  A low-density plume of neon is introduced in vacuum using a flow-calibrated orifice with a diameter of 100 $\mu$m held at a backing pressure of 60 Torr located 4 mm below the laser focus. We estimate a gas density of $3 \times 10^{14}$ cm$^{-3}$ from a steady-state Ansys-Fluent simulation of the gas expansion into vacuum \cite{AnsysF}. Five scintillating calorimeter detectors were placed around the focus. Four detectors were placed in the laser forward direction, with one oriented along the direction of laser propagation, two lying in the plane of laser polarization, and one outsize the polarization plane. The fifth detector was placed out of the forward direction and polarization plane, where no ATI electrons are expected as a control. We expect the higher-energy ATI electrons to be ejected further towards the laser forward direction \cite{Moore1995}\cite{Ekanayake2013} and preferentially ejected in the plane of laser polarization \cite{McNaught1998}\cite{Dichiara2008}. 

Each detector consisted of a 50 mm diameter, 40 mm long cylinder of long-lifetime (285 ns) scintillating plastic (Eljen Technologies EJ240) coupled to a photomultiplier tube (PMT) with a tapered voltage divider for optimal pulse linearity. The plastic scintillator and PMTs were encased in a vacuum-compatible PTFE housing that was made light-tight with colloidal graphite and aluminum foil. The plastic scintillator was chosen to decrease sensitivity to high-energy photons and a long scintillating lifetime allowed the detector to function as a calorimeter, with the output signal producing an integrated measurement of the energy of all electrons incident on the detector. The relatively large solid angle subtended by the detectors ($\sim$ 0.03 steradians) allowed several hundred ATI electrons to hit each detector, enabling relatively accurate calorimeter measurements with only a few shots at each laser intensity. Information about the shape of the energy spectrum was obtained by varying the thickness of aluminum shielding in front of each detector, which is explained further in the next section. 
        
        The scintillating detectors were calibrated by using standard pulse-height analysis techniques to measure absorbed energy spectra from two gamma radiation sources, Co-60 and Cs-137, at high operating voltage. An MeV-equivalent charge was obtained by measuring the location of the Compton edges in the acquired spectra and comparing to absorbed energy spectra calculated using G4beamline \cite{Roberts2007}, a Monte Carlo particle transport software package based on Geant4. The uncertainty in this MeV-equivalent charge is between 20\% and 25\%, and is the dominant source of uncertainty in the ATI electron energy yields. The PMT gain curves as a function of bias voltage were characterized by exposure to a Q-switched Nd:YAG laser light source.

	The output current pulse from each PMT was recorded on a Tektronix TDS5054 oscilloscope and digitally filtered to eliminate noise from electromagnetic pulse (EMP) on shot. The current pulse amplitude and integrated charge obey a linear relationship during normal detector operation. The upper saturation limit showed increasing amplitude without increasing charge, and corresponded to $~7$ GeV of integrated ATI electron energy absorbed in the scintillator. The lower charge limit corresponded to a regime where residual ringing disrupted the linear relationship, with large amplitude current pulses integrating to near-zero charge, and was chosen to prevent uncertainty induced by detector ringing from dominating over the uncertainty from the charge calibration. Measurements falling outside these limits are excluded from the figures presented in this paper but the lowest detector charge threshold is marked if appropriate.
        
        We performed a series of control tests to confirm that the observed scintillator signal was caused by high-energy electrons. We compare the signal with minimal detector shielding at the $30^{\circ}$ and $110^{\circ}$ (control) positions and found the signal at the forward detector position was at least two and a half orders of magnitude greater than the signal observed at the control detector position. We swapped the scintillating detectors between these two positions several shots after the experiment start to verify that the observed signal in the laser forward direction was not an artifact of that particular scintillating detector. Multiple control shots were taken with no target gas to confirm the signal was not just electromagnetic pulse (EMP). We also verified the effect was intensity-dependent and not energy dependent by stretching the pulse to a length of 2 ps and observing the signal to disappear. The observation of a much stronger intensity-dependent signal in the laser forward direction means that the observed signal is generated by high-energy electrons, as it cannot be attributed to a detector artifact. We also included a number of helium control shots in this study to demonstrate the contribution of any vacuum-accelerated L-shell electrons to the signal.
                
        We used the Texas Petawatt Laser in rod shot mode (64 mm Nd:silicate amplifier only) allowing an increased repetition rate of 2.5 shots per hour. We installed a \textit{f}/1.4 off-axis parabolic mirror (OAPM) to reach an intensity exceeding $2 \times 10^{20}$ W/cm$^{2}$. Laser intensity was calculated using the indirect Output Sensor Package (OSP) of the Texas Petawatt Laser, which includes diagnostics to measure near field, equivalent far field (focal spot), wavefront, pulse duration, and energy \cite{Tiwari2019}. Pulse duration was deconvolved from a second-order autocorrelation assuming a Gaussian pulse shape. Wavefront was measured using a PHASICS SID4 wavefront sensor, and a deformable mirror was used to optimize the laser wavefront before every shot. 
        
        The focal spot in the target chamber was measured using a Mitutoyo 50x plan apochromatic long-working distance microscope objective (0.55 numerical aperture), a 200 mm achromatic lens, and a vacuum-compatible CCD camera mounted in a Thorlabs optical cage system. We estimate the central maximum of the focal spot to have a full-width half-maximum (FWHM) of $2.6 \pm 0.2$ $\mu$m from Gaussian fittings of focal spot images. The far-field diagnostic plane measured at OSP does not necessarily coincide with the plane of highest laser intensity within the low-density gas jet due to defocus remaining in the wavefront after correction, which can lead to a systematic underestimate of the laser intensity. The focal spot profile used to compute the peak laser intensity was calculated from the measured wavefront including all aberrations except defocus. An inverted-field autocorrelator was used to diagnose pulse-front tilt. We estimate a 42 fs pulse front tilt from the angular shift of the far-field during grating optimization, for a total typical pulse duration of 170 fs. Intensity changes were achieved by inserting calibrated neutral density filters (ND) before the rod amplifier in the TPW laser chain. The gain of the rod amplifier remained fixed to ensure the amplified spectrum, compressed pulse duration, and laser wavefront remain the same when the pulse energy is decreased.
        
\section{ATI Model Description}

        The theoretical K-shell electron yields and energy spectra were calculated using the two-step quasi-classical models of ATI. A Monte Carlo simulation of tunneling ionization in the laser field using the ADK-PPT model of ionization\cite{Ammosov1986}\cite{Perelomov1966}\cite{Popov2004} predicted the initial conditions for K-shell electrons in the laser field. The static tunneling ionization rate for a single electron expressed in atomic units is given by: 
\begin{eqnarray}
\label{adkppt}
W_{ADK-PPT}(t) = C_{n^{*}l^{*}}^{2} I_{p} \frac{(2l + 1)(l + |m|)!}{2^{|m|} |m|! (l - |m|)!} \times  \nonumber \\
\bigg(\frac{1}{2}\tilde{F}(t)\bigg)^{1 + |m| - 2n^{*}}exp\bigg(-\frac{2}{3\tilde{F}(t)}\bigg)
\end{eqnarray}
where the reduced field strength $\tilde{F}(t)$ is defined as 
\begin{equation}
\tilde{F}(t) = \frac{\sqrt{\textbf{E}^{*}(t) \textbf{E}(t)}}{(2I_{p})^{3/2}}
\end{equation}
where $I_{p}$ is the ionization potential and $l, m$ are the orbital quantum numbers. The extension of the original PPT model by Ammosov, Krainov, and Delone introduced an effective principle number $n^{*}$ and an effective orbital quantum number $l^{*}$ given by
\begin{eqnarray}
n^{*} = \frac{Z}{\sqrt{2I_{p}}} \\
l^{*} = n^{*}_{o} - 1
\end{eqnarray}
where $Z$ is the residual charge ($Z=1$ for neutral atoms). The constants $C_{n^{*}l^{*}}^{2}$ are expressed as 
\begin{equation}
\label{prefactor}
C_{n^{*}l^{*}}^{2} \approx \bigg[\bigg(\frac{2exp(1)}{n^{*}}\bigg)^{n^{*}}\frac{1}{\sqrt{2\pi n^{*}}}\bigg]^{2}
\end{equation}
The exponential factor in Eq. \ref{adkppt} dominates the scaling of ionization probability with laser intensity, leading to an intensity threshold for the appearance of high ion charge states. Ion motion, although it has no significant effect on the ionization yield at $10^{20}$ W/cm$^{2}$, was included \cite{Yandow2019}. 

        At each timestep, Eq. \ref{adkppt} was used to predict the probability of ionization and Monte Carlo methods were used to increment the ion charge state. Non-sequential double ionization (NDSI), inelastic tunneling\cite{Kornev2003}\cite{Zon1999}\cite{Bryan2006}, collective tunneling\cite{Kornev2009}\cite{Zon1999}, and relativistic ionization effects\cite{Milosevic2002a}\cite{Milosevic2002}\cite{Mur1998}\cite{Yakaboylu2013}\cite{Klaiber2013} were excluded from the calculations. From these initial conditions, the electron trajectories were calculated by integrating the Lorentz force equations using an adaptive-timestep Runge-Kutta (RK45) numerical method. A maximum of $10^{5}$ test electrons were simulated at each intensity, originating within an isointensity boundary where ionization outside could be neglected due to the strong ionization rate dependence on intensity. We chose a series of model intensities between $3 \times 10^{19}$ W/cm$^{2}$ and $4 \times 10^{20}$ W/cm$^{2}$ to demonstrate the model behavior above and below the barrier suppression intensity of Ne$^{9+}$. Within each volume, we chose $2.5 \times 10^{5}$ initial positions for neutral ions. From these ionization events, we calculated the energy spectrum and angular distribution of at most $10^{5}$ ATI electrons. 
        
\begin{figure}[t!]
\centering
\includegraphics[width = \linewidth]{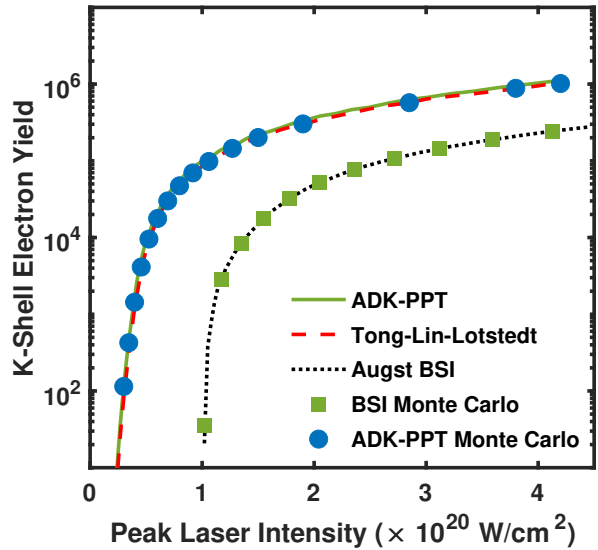}
\caption{\label{fig2} Total number of K-shell ATI electrons predicted by different models of ATI. Solid, dashed, and dotted curves are the predictions of the ADK-PPT \cite{Perelomov1966}\cite{Ammosov1986}, Tong-Lin-Lotstedt model for helium-like ions \cite{Lotstedt2020a}, and barrier suppression ionization \cite{Augst1991}. Blue circles and green squares are from Monte Carlo simulations using the ADK-PPT and BSI models, respectively. Gas density is assumed to be $3 \times 10^{14}$ cm$^{-3}$ in the laser focus. Color figures available online.}
\end{figure}

        An additional ATI Monte Carlo model using the Augst barrier suppression ionization (BSI) model \cite{Augst1991} was developed as well. The simulations were performed similarly, except the ionization event occurred at the timestep $\tilde{F} > 1/16n^{*}$ and did not occur otherwise. Figure \ref{fig2} shows the K-shell ATI electron yield predicted by the Monte Carlo simulation as a function of laser intensity assuming a gas density of $3 \times 10^{14}$ cm$^{-3}$. Analytic predictions of the ADK-PPT model, the Tong-Lin-Lotstedt model for tunneling rate near the barrier suppression regime  \cite{Tong2005}\cite{Lotstedt2020a}, and the Augst BSI model compare favorably with the Monte Carlo modeling. The effect of barrier suppression corrections on the tunneling ionization rate, which is predicted to be significant with pulses shorter than 15 fs \cite{Kostyukov2018}, can be safely neglected for this relatively long pulse duration. 
        
\begin{figure}[t!]
\begin{minipage}{\linewidth}
\subfloat{\label{prov_fig3a}\includegraphics[width = \linewidth]{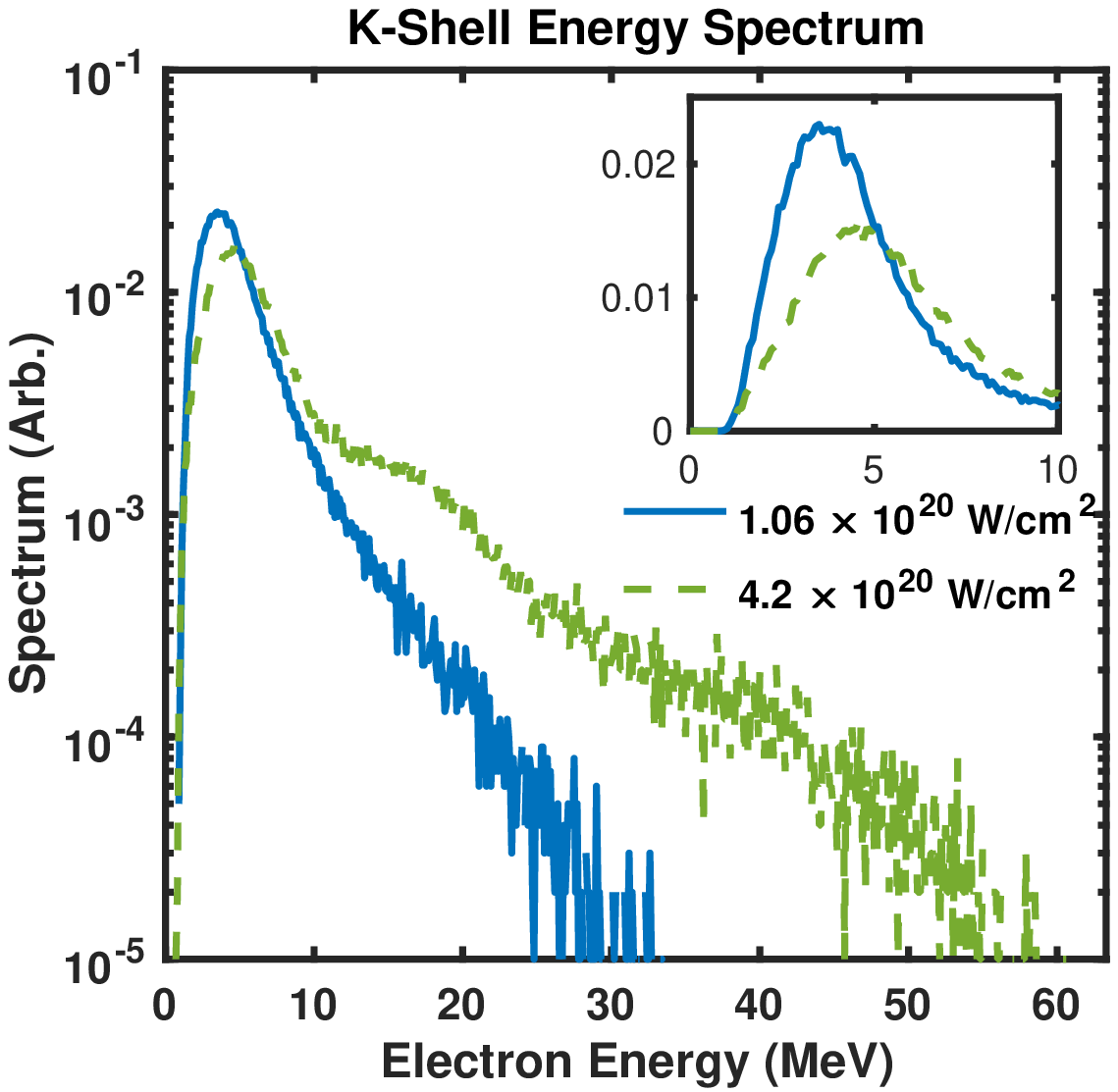}}

\subfloat{\label{prov_fig3b}\includegraphics[width = \linewidth]{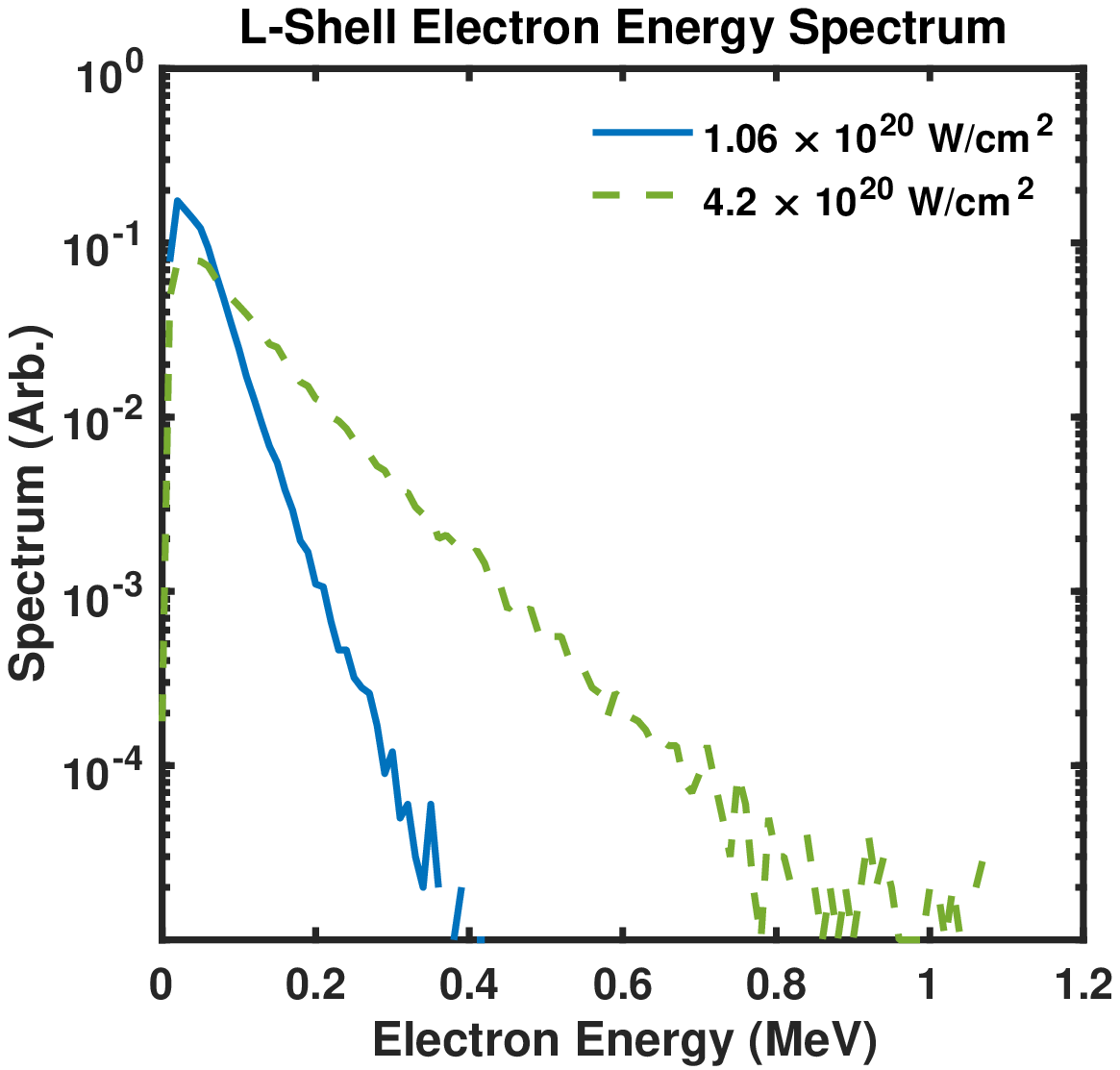}}
\end{minipage}
\caption{a) Simulated energy spectrum of the K-shell electrons at an intensity of 1.06 $\times 10^{20}$ W/cm$^{2}$ (blue) And 4.20 $\times 10^{20}$ W/cm$^{2}$ (green, dashed). Inset figures shows same curves on a linear scale. b) Simulated L-shell electron spectrum at the same two intensities. Color figures available online.}
\end{figure}
                
     The laser focal spot is computed for every shot but we lack information on the exact structure of the phase fronts as the laser pulse propagates through the focal plane, so we make a considerable number of simplifying assumptions when modeling the laser focus. Ionization rate depends strongly on intensity, so we assume the K-shell electrons are all produced in the central maximum at the focal plane, and we do not consider laser energy scattered outside the central maximum in the model. We also make the assumption that we can treat this central maximum as a Gaussian laser focus with nonparaxial corrections included up to fifth order in the diffraction angle \cite{Salamin2007}. We assume a focus with a $1/e^{2}$ diameter of 2.25 $\mu$m, which we estimated from direct measurements in the target chamber. During the experiment rod shots, we estimate the 1/e$^{2}$ spot of the central maximum was 2.2 $\pm$ 0.2 $\mu$m. We incorporate the measured pulse front tilt of 42 fs by assuming a Gaussian pulse shape with an intensity FWHM of 170 fs. Similar approaches have been taken to model the laser fields at focus in previous experimental studies  \cite{Dichiara2008}\cite{Ekanayake2013}\cite{Moore1995}. Some particle-in-cell (PIC) methods have shown promise for predicting the energy spectra of vacuum-accelerated electrons at an intensity of $10^{19}$ W/cm$^{2}$ \cite{Ivanov2018}, but no such methods have been applied to simulating ATI electron dynamics in a laser field with a more complex spatial structure than a Gaussian beam.

\begin{figure}[b!]
\centering
\includegraphics[width = \linewidth]{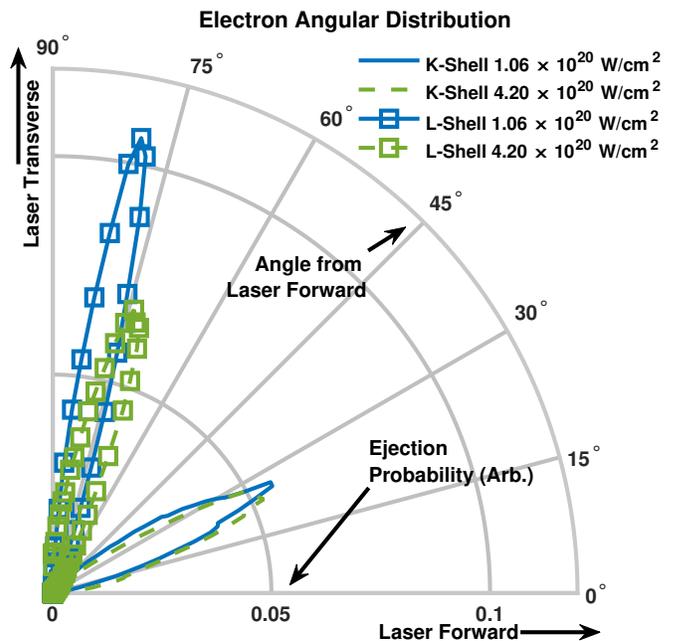}
\caption{\label{prov_fig3c} Angular distribution of the K-shell electrons (no markers) and L-shell electrons (square markets) at laser intensities of $1.06 \times 10^{20}$ W/cm$^{2}$ (solid, blue) and $4.2 \times 10^{20}$ W/cm$^{2}$ (dashed, green). Color figures available online.} 
\end{figure}

        Figures \ref{prov_fig3a} and \ref{prov_fig3b} show the energy spectra of the K-shell and L-shell ATI electrons, respectively, predicted by the Monte Carlo ADK-PPT modeling at two intensities ($1.06 \times 10^{20}$ W/cm$^{2}$ and $4.2 \times 10^{20}$ W/cm$^{2}$). The predicted angular distributions of the ATI electrons at the same intensities are shown in Figure \ref{prov_fig3c}. The modeling predicts the ATI electron energy spectra and angular distributions are be strongly modulated, with the higher-energy K-shell electrons expelled at an angle around 25$^{\circ}$ from the laser forward and the lower-energy L-shell electrons expelled an angles greater than $60^{\circ}$ from the laser forward direction.
        
	The modeled K-shell energy spectra in Figure \ref{prov_fig3a} also show that the number of high energy electrons ($> 15$ MeV) produced can increase by more than an order of magnitude as the laser intensity increases toward the maximum intensity used in the experiment. However, other features of the K-shell electrons are relatively stable, with the peak of the energy spectrum in Fig \ref{prov_fig3a} (inset) increasing from 3.5 MeV to 4.7 MeV and the angular distribution in Figure \ref{prov_fig3c} nearly unchanged. The modeling predicts that the energy yield attributed to the highest-energy electrons, which are observed on the $0^{\circ}$ detector, demonstrate a stronger scaling with laser intensity than the other detectors. The energy yield at this position increases with intensity due to both the larger number of electrons generated in the focus, as shown in Figure \ref{fig2}, and the larger number of electrons in the high-energy tail of the spectrum. The energy yields detected at the $30^{\circ}$ position, where the electron energies and angular distribution change little with increasing intensity, will display a scaling dominated by the total number of electrons produced by the K-shell ionization.

        The L-shell electrons, shown in Figure \ref{prov_fig3b}, are predicted to obtain energy less than 1 MeV and be ejected from the focus at an angle around $70^{\circ}$, and can therefore be filtered from the K-shell electrons in energy and angle, but we must treat these model predictions with caution. Vacuum acceleration of electrons demonstrates very strong sensitivity to initial position in the laser focus \cite{Popov2008}, leading to the possibility that the simulation method may under-sample initial positions in the focus that yield L-shell electrons that are accelerated to higher energies or ejected further towards the laser propagation direction. The last study of vacuum acceleration of electrons from ionized helium in this intensity regime ($\sim 3 \times 10^{20}$ W/cm$^{2}$) found disagreement between the measured angular distribution of vacuum-accelerated electrons and the angular electron distributions predicted by particle-in-cell modeling. The authors suggested this discrepancy may be caused by poor sampling of the focal volume in their simulations \cite{Kalashnikov2015}, and there is evidence of L-shell electrons scattered as far forward as $30^{\circ}$ from the laser propagation direction from the helium control shots. 
            
\begin{figure}[t!]
\centering
\includegraphics[width = \linewidth]{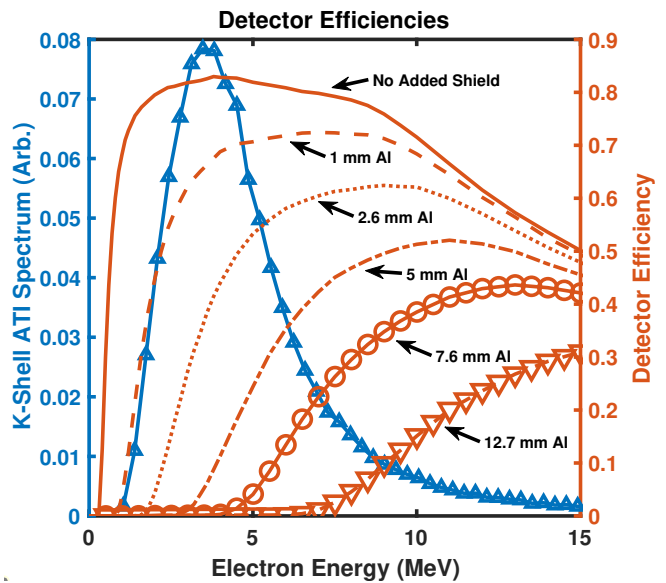}
\caption{\label{fig4} Detector efficiencies at different aluminum thicknesses (right axis) alongside a simulated K-shell ATI electron spectrum using the ADK-PPT model at intensity of $1.06 \times 10^{20}$ W/cm$^{2}$ (left axis). Color figures available online.}
\end{figure}

        The simulated ATI electron yields and energy spectra incident on each detector are used to calculate the energy deposited in the plastic scintillator. Several thicknesses of aluminum shielding were used in the experiment to block lower-energy electrons. The detector efficiencies for each shielding thickness were calculated using G4beamline \cite{Roberts2007}, a Monte Carlo particle transport software package based on Geant4, that includes energy deposited in the plastic scintillator by electrons, positrons, and high-energy photons generated in the interaction. The detector efficiencies are shown in Figure \ref{fig4}. The detector efficiency at each energy and shielding thickness is calculated from the simulated visible energy deposited in the scintillating plastic by a monoenergetic beam of $10^{4}$ electrons with a divergence similar to the incident ATI electrons.
        
        The predicted energy yield in the scintillators can be computed by combining the electron yields and energy spectra from the ATI modeling with the calculated detector efficiencies. The predicted ATI electron energy yield is given by 
\begin{equation}
Y_{ATI}(\theta, \phi, Z) = \int_{0}^{E_{max}} w_{V}E'p(E', \theta, \phi)\eta_{Al}(E', Z)dE'
\end{equation}
where $p(E, \theta, \phi)$ is the energy spectrum (count) at the detector position, $\eta_{Al}(E, Z)$ is the detector efficiency with an aluminum shielding thickness of Z, and $w_{V}$ is a volume weighting factor corresponding to the number of real K-shell electrons produced per simulated ATI electron. The volume weighting factor is calculated by using a gas density of $3 \times 10^{14}$ cm$^{-3}$ and a confocal volume estimated by integrating a Gaussian beam volume bounded by the same isointensity shell used in the model simulations. 

        We can gain some information about the energy spectrum of electrons at a given intensity by varying the shielding thickness $Z$ and measured the energy deposited. Although the energy integration cannot be uniquely inverted to give an electron spectrum, we can compare the predicted energy deposited to the observed energy deposited and search for energy ranges where the ATI model spectrum either overestimates or underestimates the electron number. From the efficiency curve shapes, we conclude this method has poor resolution for electron energies above $10$ MeV but can yield some energy information in the 0.3-6 MeV energy range. 
\section{ATI Electron Energy Yields}
\label{section4}

A number of laser intensity scans were performed using different shielding configurations. On all unshielded detectors in the laser forward direction the integrated electron energy increase up to almost three orders of magnitude. Helium control shots show that some of the electrons accelerated toward these detectors are low-energy L-shell electrons ($<$ 1.4 MeV) scattered in the laser forward direction by vacuum acceleration. 

\begin{figure}[b!]
\centering
\includegraphics[width = \linewidth]{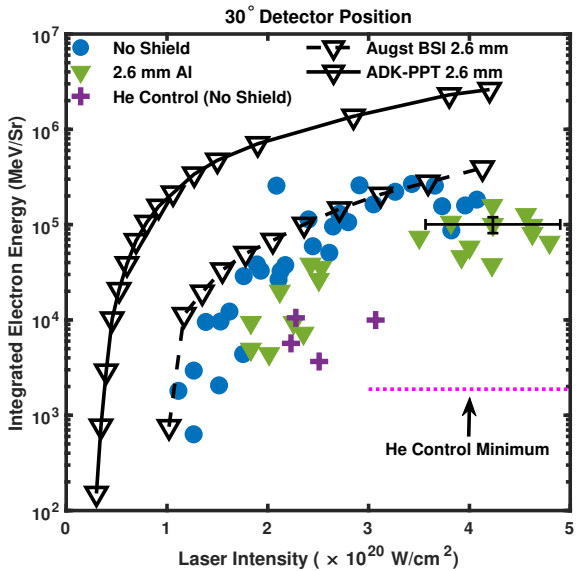}
\caption{\label{ch7fig6} Measured electron energy deposited in a scintillating detector located at $30^{\circ}$ from the laser forward direction in the polarization plane. Unshielded and shielded (2.6 mm aluminum) intensity scans are given by blue circles and green triangles, respectively. Helium control shots in the unshielded configuration are given by purple crosses and the detector charge floor for helium control shots in the shielded configuration is marked. Monte Carlo simulations using ADK-PPT (solid) and BSI (dashed) models shown for comparison. Color figures available online.}
\end{figure}

Figure \ref{ch7fig6} shows the energy absorbed in the scintillating detector at $30^{\circ}$, where the number of K-shell ATI electrons is expected to be the highest. Intensity scans with the unshielded scintillator (blue circle) and a shielded configuration (green triangles) are shown to compare the total integrated electron energy with the integrated energy from electrons with energy greater than 2.8 MeV. Helium control shots (purple crosses) demonstrate that the L-shell electrons contribute some of the deposited energy in the unshielded configuration. The helium control shot energy yield is about an order of magnitude lower than the neon yield at $2.5 \times 10^{20}$ W/cm$^{2}$, demonstrating that the neon L-shell electrons account for $\sim 1/2$ of the observed energy yield when accounting for the difference in electron density at the focus. Two helium control shots taken in the shielded configuration with the same backing pressure ($n_{a} \sim 3 \times 10^{14}$ cm$^{-3}$ yielded no repeatable signal, with the dynamic range floor for these control shots marked on Figure \ref{ch7fig6}. The helium control shots establish an upper limit of 2.8 MeV for vacuum-accelerated, which is slightly lower than the maximum energy of vacuum-accelerated electrons by Kalashnikov near this angle in this intensity regime \cite{Kalashnikov2015}.

The shielded measurements show a threshold intensity around $2\times 10^{20}$ W/cm$^{2}$, above which the probability of electron production with energy $>$ 2.8 MeV increases rapidly with intensity. A scaling transition around $3\times 10^{20}$ W/cm$^{2}$ marks the saturation intensity where the scaling transitions from an ionization probability scaling dominated by the exponential term in Eq. \ref{adkppt} to a focal volume scaling. The threshold-like behavior and scaling transition are features of ATI that are mirrored both Monte Carlo models, although neither model correctly predicts the threshold or saturation intensities and both overestimate the ionization yield.

\begin{figure}[t!]
\centering
\includegraphics[width = \linewidth]{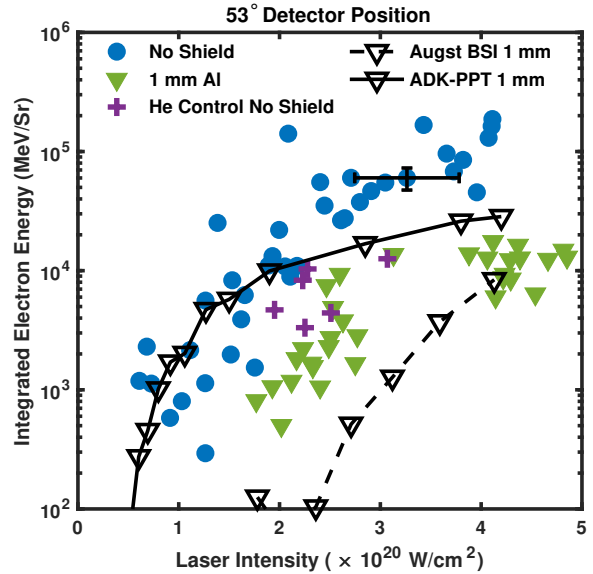}
\caption{\label{ch7fig10} Measured energy deposited in a scintillating detector located at $53^{\circ}$ from the laser forward direction in the polarization plane. Unshielded and shielded (1 mm aluminum) intensity scans are given by blue circles and green triangles, respectively. Helium control shots in the unshielded configuration are given by purple crosses. Monte Carlo simulations using ADK-PPT (solid) and BSI (dashed) models shown for comparison. Color figures available online.}
\end{figure}

\begin{figure}[b!]
\includegraphics[width = \linewidth]{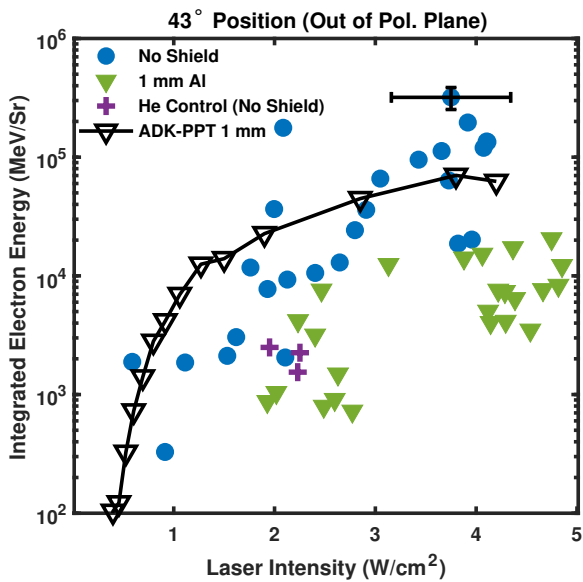}
\caption{\label{ch7fig11} Measured energy deposited in a scintillating detector located at $43^{\circ}$ from the laser forward direction in the polarization plane. Unshielded and shielded (1 mm aluminum) intensity scans are given by blue circles and green triangles, respectively. Helium control shots in the unshielded configuration are given by purple crosses. Monte Carlo simulations using the ADK-PPT ionization model (solid) shown for comparison. Color figures available online.}
\end{figure}

Similar laser intensity scans at two additional positions are presented in Figures \ref{ch7fig10} and \ref{ch7fig11}, corresponding to positions $53^{\circ}$ from the laser forward direction (in polarization plane) and $43^{\circ}$ from the laser forward direction ($60^{\circ}$ out of the polarization plane), respectively. These positions were chosen due to space restrictions in the target chamber and experimental setup. The helium control shots with no shielding installed are comparable to the deposited energies measured with neon in both cases, showing the L-shell electrons will contribute to the signal substantially. Installing a 1 mm aluminum shield, which blocks electrons with energy $<1.4$ MeV, decreases the electron energy yields an order of magnitude at each detector. The electron energy yields in the shielded configuration show the same characteristic ATI features, the threshold and saturation intensities, seen in \ref{ch7fig6}, with the saturation effect somewhat more exaggerated because the K-shell ATI electrons will be ejected further forward in the laser direction as laser intensity continues to increase. The unshielded measurements dominated by lower-energy electrons do not display the clear scaling change visible in the shielded measurements, and so they are likely L-shell electrons scattered in the laser forward direction by a vacuum acceleration process.

At these larger angles, the ADK-PPT simulations tended to overestimate the electron energy yields while the Augst BSI model tended to be an underestimate, instead predicting a greater proportion of higher-energy ATI electrons that would scatter further forward in the focus. The BSI model also exhibited an unexpectedly strong polarization dependence for low-energy ATI electrons because the probability of being ``born'' into the field off a laser cycle peak is higher for ATI electrons produced by the rising edge of the laser focus and scattered out before the arrival of peak laser intensity. An insufficient number of test electrons in the BSI simulations were scattered toward the $43^{\circ}$, so only the ADK-PPT model is shown in Fig. \ref{ch7fig11}. 

\begin{figure}[t!]
\centering
\includegraphics[width = \linewidth]{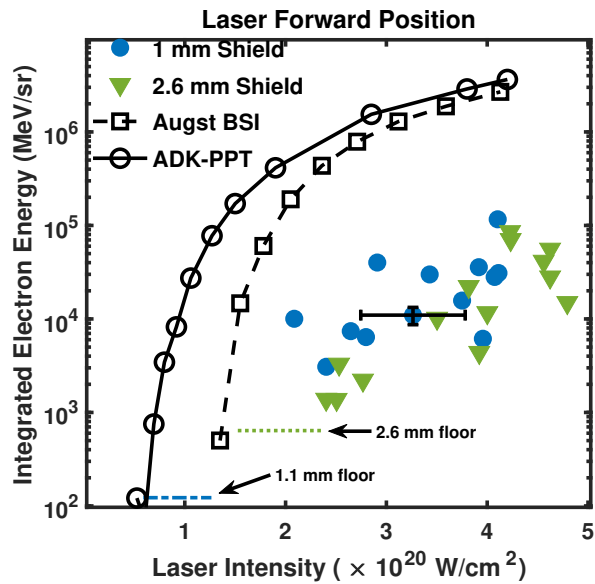}
\caption{\label{ch7fig8} Measured electron energy deposited in a scintillating detector located at on the laser propagation axis. Two shielded (1 mm and 2.6 mm aluminum) intensity scans are given by blue circles and green triangles, respectively. Lowest detector charge floors are marked over the intensity range where shots were taken in each shielding configuration. Monte Carlo simulations using ADK-PPT (solid) and BSI (dashed) models shown for comparison. Color figures available online.}
\end{figure}

Figure \ref{ch7fig8} shows the measured electron energy deposited in the scintillating detector oriented in the laser forward direction, shielded with a minimum of 1 mm of aluminum to block electrons with energy $<$ 1.4 MeV. We observe a threshold appearance intensity of $2 \times 10^{20}$ W/cm$^{2}$ for high-energy electrons in the laser forward direction, which are found to be in the 10-16 MeV range \cite{YandowPRL2023}. The measured ATI electron energy yields along the laser forward direction fall nearly two orders of magnitude lower than the ADK-PPT and BSI simulation predictions. While a scaling transition is not obvious in the measurements, it is important to note that the average energy of these electrons is much higher than at other detector positions. A single 15 MeV electron incident on this detector would yield a $\sim 500$ MeV/Sr response, so some of these measurements between 2-3 $\times 10^{20}$ W/cm$^{2}$ represent a single-digit number of electrons, and uncertainty due to sampling statistics obscures the scaling transition. Measurements falling below the instrument dynamic range floor (hollow markers) at $10^{20}$ W/cm$^{2}$ show that not even a single one of these ATI electrons exceeding 10 MeV was detected below the threshold intensity. 

We do not observe good quantitative agreement between the predicted ATI energy yields of either Monte Carlo model and the measured energy yields, although the measurements demonstrate self-consistent qualitative features of tunneling ionization between the four detector positions. All show an appearance intensity for a population of high-energy electrons above $2 \times 10^{20}$ W/cm$^{2}$ and three of the four detector positions show a consistent saturation intensity. The ADK-PPT tunneling ATI model predicts these features will appear on all detectors at about the same intensity, although the model intensity underestimates the experimental intensity by a factor of 3-4. The BSI ATI model predicts a narrower angular distribution of ATI electrons that broadens as the intensity increases, the focal volume grows, and a broader range of electron initial conditions over the focal volume and pulse duration are sampled. This broadening of the ATI electron angular distribution as intensity increases leads to the higher predicted appearance intensity at 53$^{\circ}$. Therefore, the measured ATI energy yields are more consistent with some form of tunneling process that allows for electrons to originate from a wider range of initial conditions below the saturation intensity than it is with a true intensity threshold process.

\section{Electron Energies}
The limited number of laser shots and the low density of target gas necessary to avoid collective plasma effects prevented measurement of the electron spectrum using a magnetic spectrometer. We placed a series of aluminum filters of different thicknesses in front of the scintillating plastic to gain spectral information. While such a method provides only crude information about the energy spectrum, it can be used to show the maximum ATI electron energy is between 10-16 MeV. A comparison to the maximum energies predicted by ATI models shows the maximum ATI energy range consistent with the measurements falls between the predictions of relativistic and nonrelativistic ponderomotive models \cite{YandowPRL2023}. 

\begin{figure}[t!]
\begin{minipage}{\linewidth}
\label{fig9}
\subfloat{\label{fig9a}\includegraphics[width = 1\linewidth]{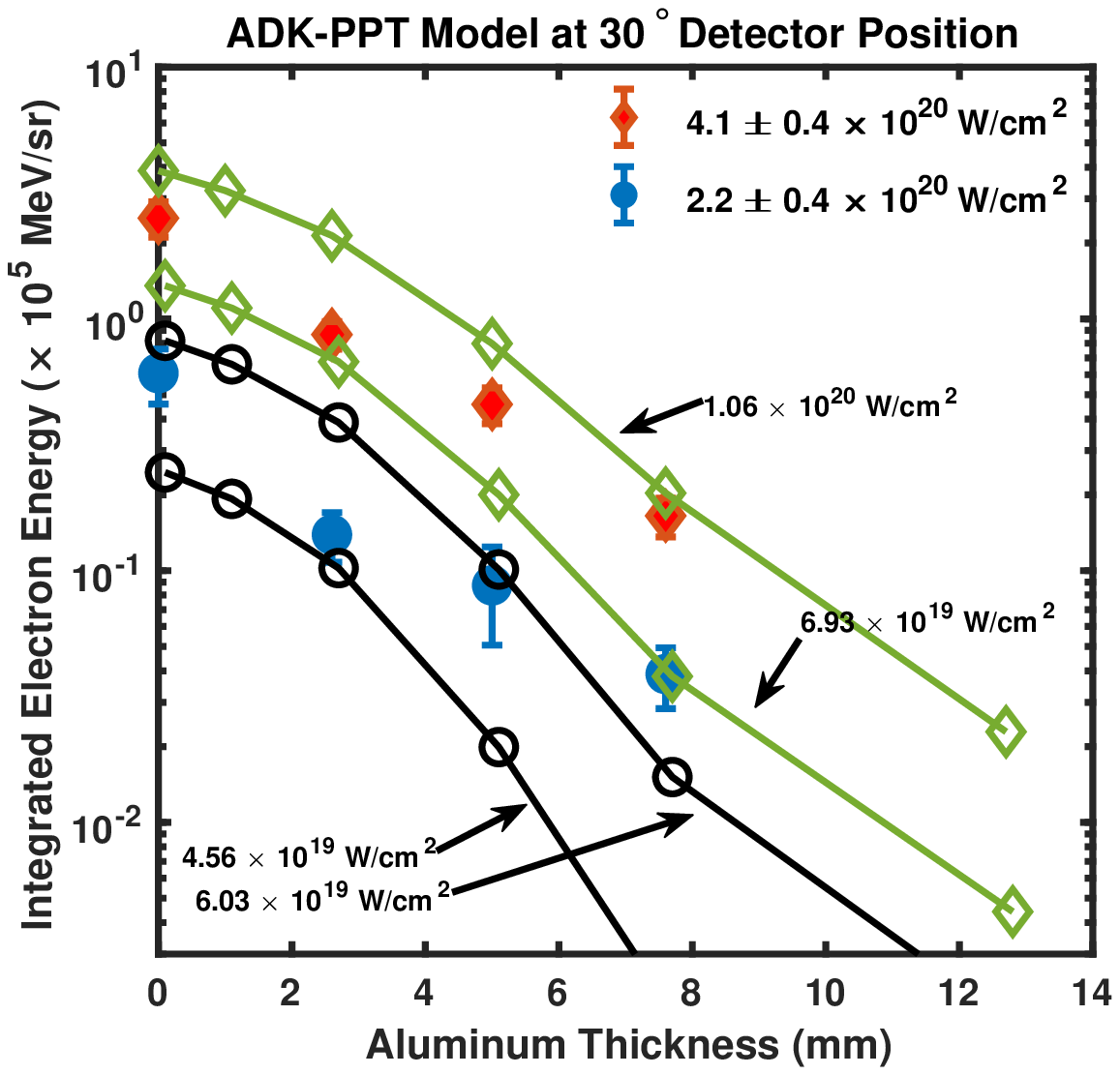}}

\subfloat{\label{fig9b}\includegraphics[width = 1\linewidth]{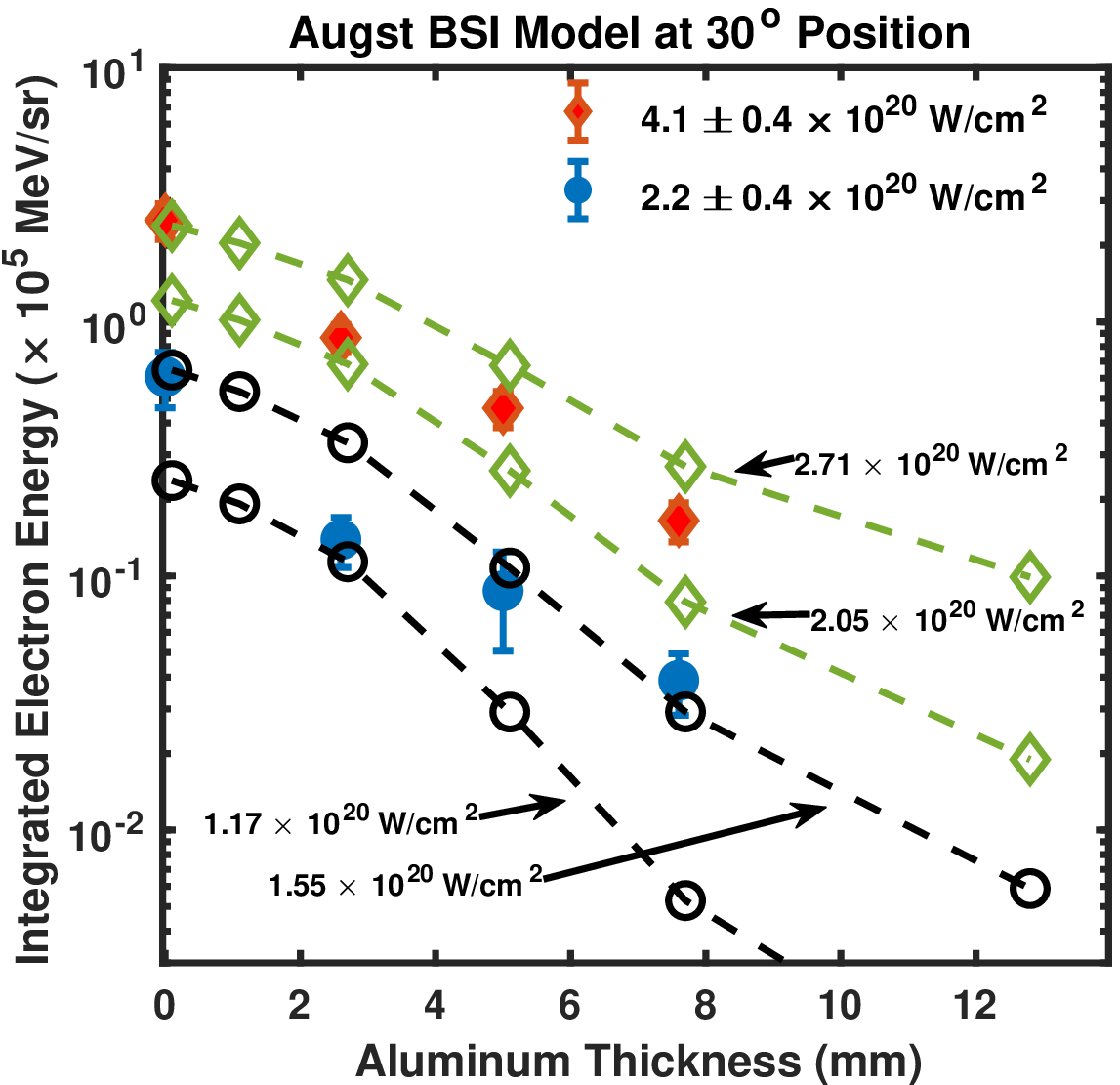}}
\end{minipage}
\caption{Electron energy yields measured at the $30^{\circ}$ position with varying shield thicknesses at two average intensities, $4.1 \pm 0.4 \times 10^{20}$ W/cm$^{2}$ and $2.2 \pm 0.4 \times 10^{20}$ W/cm$^{2}$ compared to closest predictions of a) the ADK-PPT Monte Carlo model (solid lines) and b) the Augst BSI Monte Carlo model(dashed lines). Color is added for clarity. Color figures available online.}
\end{figure}

\begin{figure}[t!]
\centering
\includegraphics[width = \linewidth]{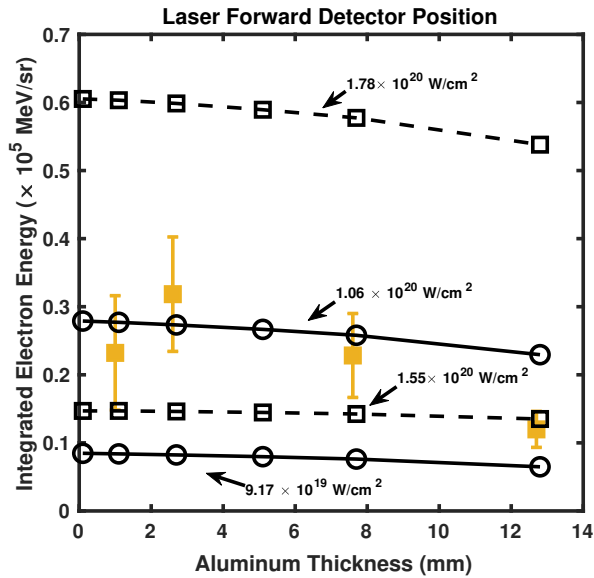}
\caption{\label{fig10} Electron energy yields measured along the laser forward direction with varying shield thicknesses at an average intensity of $4.1 \pm 0.4 \times 10^{20}$ W/cm$^{2}$. Solid curves (open circles) are ADK-PPT model predictions and dashed curved (open squares) are BSI model predictions.}
\end{figure}

Figures \ref{fig9a} and \ref{fig9b} show the measurements of integrated electron energy at the $30^{\circ}$ positions at two average laser intensities. The predictions of ADK-PPT Monte Carlo model and BSI model at several intensities are marked on Figures \ref{fig9a} and \ref{fig9b}, respectively.  Figure \ref{fig10} shows the measured electron energy yield along the laser forward direction, and the predictions of the ADK-PPT (solid) and BSI (dashed) models, and is consistent with a maximum ATI electron energy in the 10-16 MeV range \cite{YandowPRL2023}. 

Both models only show that quantitative agreement with the electron energy yield measurements is only possible when the laser model intensity is taken to be significantly less than the laser intensity computed using indirect laser diagnostics. As with the laser intensity scans discussed in Section \ref{section4}, we observe the ADK-PPT ATI model provides a more consistent description of the measurements between different detector positions, even though the model intensity is four times lower than the estimated laser intensity in the experiment. The BSI model does not make predictions that are consistent between the on-axis and $30^{\circ}$ detectors, with the intensity that is most consistent with the electron energy yields for the on-axis detector in Figure \ref{fig10} ($1.55 \times 10^{20}$ W/cm$^{2}$) underestimating the measurements at $30^{\circ}$ by a factor of $\sim 5$. The ADK-PPT model shows a more consistent model intensity around $10^{20}$ W/cm$^{2}$ between the two detector positions. 

Some qualitative statements about the shape of the spectrum can be gathered by comparing the measurements in Figure \ref{fig9a} to the detector efficiency curves in Figure \ref{fig4}. We cannot draw many conclusions about the unshielded measurement because of the evidence of forward-scattered L-shell electrons shown by the control shots in Figure \ref{ch7fig6}, so we cannot make a statement about the population of electrons with energy $< 2.8$ MeV. The Monte Carlo ADK-PPT model predicts a steeper electron energy yield drop-off than the measurements, corresponding to a model overestimate of the proportion of electrons with energy between 2.8-4.7 MeV and an underestimate of the number of electrons with energy $> 6.5$ MeV. The BSI model predictions in Figure \ref{fig9b} show a decrease in electron energy yield with shield thickness that is more consistent with measurements, which could indicate an ionization process with higher onset intensity than predicted by the ADK-PPT model. 

The electron energy yields predicted by the modeling at the on-axis position shown in Figure \ref{fig10} should be interpreted with care because the Gaussian focus assumption made in the model is not a realistic description of the laser fields. Higher-order spatial modes experience increased Guoy phase shifts as the beam passes through the focus, which will limit the distance a relativistic electron can stay in phase with the peak of the paraxial laser electric field to a fraction of a Rayleigh range, which should substantially decrease the maximum ATI electron energy. The fields of higher-order spatial modes will also scatter high-energy electrons over a larger range of angles than expected from a Gaussian model, which could explain why the ADK-PPT model underestimates the electron energy yield at the $30^{\circ}$ detector with 7.6 mm of shielding in Figure \ref{fig9a}. Further development of ATI simulation techniques to incorporate a more realistic model of the laser fields is necessary to further study ATI electrons and develop laser intensity diagnostics using ATI electrons. 

We performed a similar analysis for the detectors at the $53^{\circ}$ and $43^{\circ}$, and found that the installation of 1 mm aluminum shield decreased the energy deposited more than an order of magnitude as seen in Figs. \ref{ch7fig10} and \ref{ch7fig11}. No repeatable signal was observed when 2.6 mm of shielding was used, limiting the maximum ATI electron energy to below 2.8 MeV at these two angles. 

\section{Conclusion}
We report the first observation of ATI electrons with energies exceeding 10 MeV as well as the first indirect evidence of the ionization of helium-like neon in an intense laser field to the best of our knowledge. We measured the energy deposited in an array of scintillating detectors by high-energy ATI electrons and performed scans of laser intensity in several configurations and presented a comparison with the two Monte Carlo models of neon K-shell ionization.The ADK-PPT ATI model predicted roughly consistent appearance and saturation intensities between the four detector positions, a qualitative prediction consistent with the experimental measurements, although the ADK-PPT model significantly underestimated these intensities. These qualitative features were not predicted in the BSI Monte Carlo modeling because a probabilistic tunneling process allows for a much less restricted range of electron initial positions in the focal volume and laser phases at ionization.

ADK-PPT model-derived intensities derived from ionization yield measurements in prior studies have not always demonstrated consistency with laser intensity calculated from indirect diagnostic measurements or self-consistency when different atomic species are used. Ionization of lithium-like argon (Ar$^{16+}$) has been demonstrated to occur in an intensity range from $1-2 \times 10^{19}$ W/cm$^{2}$ in two different studies \cite{Chowdhury2001}\cite{Yamakawa2003}. Ionization yields of xenon in the same laser field were found to give an ADK-PPT model intensity of $3.5 \times 10^{18}$ W/cm$^{2}$, much lower than the indirectly estimated intensity of $2.6 \times 10^{19}$ W/cm$^{2}$ or argon-yield ADK-PPT model-derived intensity of $1.3 \times 10^{19}$ W/cm$^{2}$ \cite{Yamakawa2003}. The authors emphasized the repeatability of their results but were not able to provide a theoretical explanation for the systematic decrease of model-derived intensity with atomic number. Chowdhury \textit{et al}. similarly calculated a model intensity from precision measurements of argon charge states and found a similarly low model intensity, although it was within the lower bound of their experimental intensity uncertainty \cite{Chowdhury2001}. An ADK-PPT model intensity shift factor of $\sim 4$ was not expected for ionization of helium-like neon given the simplicity of the electronic shell structure and given how precisely helium ionization yields agree with the ADK-PPT model \cite{Walker1994}.

Some recent modifications to the ADK-PPT model have been proposed to account for barrier suppression effects for helium-like ions \cite{Lotstedt2020a}, but they are typically more relevant for pulses much shorter than 170 fs  \cite{Kostyukov2018} and L-shell or M-shell orbitals \cite{Ciappina2020}, which we confirmed in the calculations presented in Figure \ref{fig2}. Relativistic corrections that suppress the ionization rate are predicted to be negligible at an intensity of 10$^{20}$ W/cm$^{2}$ \cite{Milosevic2002a}\cite{Milosevic2002}. The spectral information we were able to obtain by increasing the shielding thickness at $30^{\circ}$ may be consistent with a higher threshold intensity accelerating electrons ejected at this angle to higher energies but the model of the laser fields is not realistic enough to demonstrate this agreement conclusively. Our observation of a neon K-shell ionization intensity above $10^{20}$ W/cm$^{2}$ may be a reason why it has not been reported in previous studies, but no study has explicitly stated that neon charge states were not observed in this intensity range. Momentum conservation during the ionization process will accelerate the ions to energies on the order of tens of eV, so spectrometer design in previous studies may have been a factor as well.

Our observation of forward-scattered L-shell electrons is unexpected from the simplified model of the laser focus used in this paper but is consistent with other experiments reported. Kalashnikov \textit{et al}. report vacuum accelerated electrons from helium over a similar laser intensity range and angular distribution \cite{Kalashnikov2015}. They also found disagreement with the angular distributions of vacuum-accelerated electrons predicted by both their particle-in-cell modeling and analytical calculations, which predicted a local maximum around $20^{\circ}$ for forward-scattered electrons. Instead they observed the electron number to increase monotonically as angle increased from 5$^{\circ}$ to 70$^{\circ}$, which they attribute to poor sampling of initial conditions in the focal volume. A comprehensive model of the L-shell electrons in the detected energy range ($>$ 0.3 MeV) will likely have to into account pulse shape  \cite{Vais2018}, focal spot asymmetry \cite{Hegelich2023}, and a more realistic model of laser fields at the focal plane to match experiment. 

At laser intensity exceeding 10$^{21}$ W/cm$^{2}$, ATI electrons from the K-shell of argon ($> 3 \times 10^{21}$ W/cm$^{2}$ and krypton ($> 10^{23} $ W/cm$^{2}$) are predicted to exceed energies of 100 MeV and 1 GeV, respectively. These ATI electrons will be ejected very nearly in the laser forward direction and hold promise as a low-dose ultrafast radiation source and as a direct laser intensity diagnostic. Measurement of the ATI electron spectrum would be more straightforward than the measurements presented in this paper, as the high energy and low electron divergence would enable the use of a large-aperture magnetic spectrometer located outside the vacuum chamber and along the laser forward direction. Similar scintillating detectors could be placed behind the magnet to detect ATI electrons in different energy ranges. Vacuum acceleration of the L-shell electrons to comparable energies can be suppressed by engineering a $\sim 10^{-2}$ pre-pulse that arrives a few pulse durations before the main laser pulse \cite{Vais2018}.

\begin{acknowledgments}
A. Y. acknowledges helpful conversations with E. Chowdhury regarding the design of this experiment. This work was supported by the DOE, Office of Science, Fusion Energy Sciences under Contract No. DE-SC0021125: LaserNetUS: A Proposal to Advance North America's First High Intensity Laser Research Network, the Air Force Office of Scientific Research through Awards No. FA9550-14-1-0045 and No. FA9550-17-1-0264, and the National Nuclear Security Agency (NNSA) through Award No. NA0002008. This work was also performed under the auspices of the U.S. Department of Energy by Lawrence Livermore National Laboratory under Contract DE-AC52-07NA27344. A. Y. gratefully acknowledges the generous support of the Jane and Michael Downer Fellowship in Laser Physics in Memory of Glenn Bryan Focht.
\end{acknowledgments}

\nocite{*}
\bibliography{archival_paper_vFINAL.bib}  

\end{document}